\documentclass[preprint,amsmath,amssymb]{revtex4}
\usepackage{graphicx}
\usepackage{dcolumn}
\usepackage{bm}
\usepackage{amsmath}
\usepackage{mathrsfs}
\begin{document}
\title{Casimir effect of two conducting parallel plates in a general weak gravitational field  \\}
\author{ Borzoo Nazari  \footnote{borzoo.nazari@ut.ac.ir, corresponding author}
 }
\address{Faculty of Engineering Science, University of Tehran, Tehran,
Iran, P.O.Box: 11155-4563}

\begin{abstract}
We calculate the finite vacuum energy density of the scalar and electromagnetic fields inside a Casimir apparatus made up of two conducting parallel plates in a general weak gravitational field. The metric of the weak gravitational field has a small deviation from flat spacetime inside the apparatus and we find it by expanding the metric in terms of small parameters of the weak background. We show that the found metric can be transformed via a gauge transformation to the Fermi metric. We solve the Klein-Gordon equation exactly and find mode frequencies in Fermi spacetime. Using the fact that the electromagnetic field can be represented by two scalar fields in the Fermi spacetime, we find general formulas for the energy density and mode frequencies of the electromagnetic field. Some well-known weak backgrounds are examined and consistency of the results with the literature is shown.
\end{abstract}
\maketitle
\section{Introduction}
The quantum vacuum is a fundamental concept in theoretical physics and its properties has been widely investigated in the literature of quantum gravity and string theory. The theory of quantum fields in curved spacetime which is believed to be the low energy limit of the ultimate theory of quantum gravity, has predicted famous quantum effects in the presence of gravity. In general, due to the lack of the global symmetries in the spacetime manifold, quantum effect considerations in curved spacetime are mainly limited to the analysis of the local quantities such as the vacuum expectation value of the energy-momentum tensor, i.e. $<|T_{\mu\nu}(x)|>$, in some point $x$. In fact, the most famous results of the semi-classical theory of gravity like the Hawking radiation and the particle production in the expanding universe has been achieved from the analysis of the $<|T_{\mu\nu}|>$ in the related curved backgrounds. The most famous vacuum state effect is the Casimir effect.
An important aspect in the researches around the Casimir effect in curved spacetime is that the characteristics of the vacuum state are apparently dependent to the geometry of the background spacetime. We are also motivated to see explicitly such a dependency in this paper.
Furthermore, computation of the energy, i.e. $<|T_{00}|>$, has also been done for large number of problems in various spacetimes and sometimes \cite{Milton} has helped us to confirm the validity of the principle of correspondence in the context of the Casimir effect. Finding the total gravitational force on a set of two conducting Casimir plates \cite{Milton1} is a typical example. So, due to the importance of the stress-energy tensor, we will consider the $<|T_{\mu\nu}|>$ for the plates in a general weak background.

The Casimir effect arises when there is a boundary in our problem and it predicts a force between two uncharged conducting metals in the presence of a quantum field. The effect has been measured to a great accuracy \cite{Experiment}. We use the zero-point energy approach here although it is possible to find the Casimir force and energy without any reference to the zero-point energy \cite{Jaffe}. We may also have the Casimir effect without having a boundary at all. In fact, some non-trivial topologies in curved spacetime do the same job as a boundary does\cite{DeWitt},\cite{Ford}. The Casimir energy in curved spacetime has been also analysed by many authors (\cite{Ford,Bezerra,SorgeNew,Muniz1,Muniz2,Geyer, Bordagbook,Bimonte,Sorge,Milton,Esposito,Napolitano,NouriNazari,NouriNazari2,Dowker} and references there in).
Recently, a Casimir apparatus consisting of two ideal conducting parallel plates in the weak field limit of the Kerr and the Horava-Lifshitz spacetimes has been studied in \cite{Bezerra,SorgeNew,Muniz1,Muniz2}.  A purpose of this paper is to generalize the above analysis for scalar field doing exact solution of the Klein-Gordon equation in a general weak gravitational field. Also we extend the method for the case the electromagnetic field is present inside the plates.

The organization of the paper is as follows. In section II using the fact that the apparatus has composed of tiny pales, we will find the small deviations of the metric from flat spacetime inside the apparatus. In fact we will expand the metric up to first order in terms of the parameters of the general weak gravitational field. In section III the Klein-Gordon equation will be solved exactly inside the apparatus using the metric obtained in the previous section. In section IV mode frequencies inside the apparatus will be obtained under the influence of both Dirichlet and Neumann boundary conditions for the scalar field. Generalization to the case where the electromagnetic field is present will be done using an interesting property of the Fermi spacetime in section V. Computation of the energy momentum tensor for the scalar field for both Neumann and Dirichlet boundary conditions are done in section VI. Also the electromagnetic energy density is obtained in this section. Well known weak gravitational fields are examined in section VII. The electromagnetic energy density for the far field limit of the Kerr spacetime and the Horava-Lifschitz theory of gravity are of special interest. The final section is devoted to the conclusion.

\section{Transformation of the metric of a weak gravitational field into the Fermi metric}
In the 1+3 formalism of general relativity, stationary spacetime metric is defined by \footnote{see \cite{Lynden-Bell} for more discussions on the gravitomagnetic effects in the stationary spacetimes. Also see \cite{Landau} }
\begin{eqnarray}\label{eq01}
ds^2 = g_{00}(dx^0 -{A_i}dx^i)^2 -dl^2,
\end{eqnarray}
where $A_i=-\frac{g_{0i}}{g_{00}}$ is the so called gravitomagnetic potential and
\begin{eqnarray}\label{eq02}
dl^2 = \gamma_{ij}dx^i dx^j = (-g_{ij} + \frac{g_{0i}g_{0j}}{g_{00}})dx^i dx^j \;\;\; i, j = 1,2,3. ,
\end{eqnarray}
In the weak field slowly rotating limit ($\Phi<<1,v<<c$), the metric (\ref{eq01}) is equivalent to
\begin{eqnarray}\label{eq03}
ds^2 \approx (1 + \frac{2\Phi}{c^2} - \frac{2 {\bf A}. {\bf v}}{c^2})c^2 dt^2-(1-\frac{2\Phi}{c^2})\delta_{ij}dx^i dx^j,
\end{eqnarray}
The explicit form of a general weak gravitational field line element is as follows (see (19.13) in \cite{Misner})
\begin{eqnarray}\label{eq04}
\begin{split}
ds^2= &[1-\frac{2M}{r}+\frac{2M^2}{r^2}+O(\frac{1}{r^3})]dt^2-[4\epsilon_{ijk}S^j\frac{x^k}{r^3}+O(\frac{1}{r^3})]dx^idt \\
&-\{(1+\frac{2M}{r}+\frac{3M^2}{2r^2})\delta_{ij}+gravitational\;radiation\; terms\;that\;die\; out\; as\; O(\frac{1}{r}) \}dx^i dx^j,
\end{split}
\end{eqnarray}
Comparison between (\ref{eq03}) and (\ref{eq04}) shows that $\Phi=-\frac{GM}{r}$ is the newtonian potential, $A^i=\epsilon_{ijk}S^j\frac{x^k}{r^3}$ is the gravitomagnetic potential and ${\rm v}^i = \frac {dx^i}{dt}$. We need to have an isotropic coordinate representation of the metric in the next sections and equations (\ref{eq01}),(\ref{eq02}) and (\ref{eq03}) shows us the way we can construct an isotropic coordinate representation of any weak gravitational field.
\begin{figure}
 \includegraphics[width=0.5\linewidth]{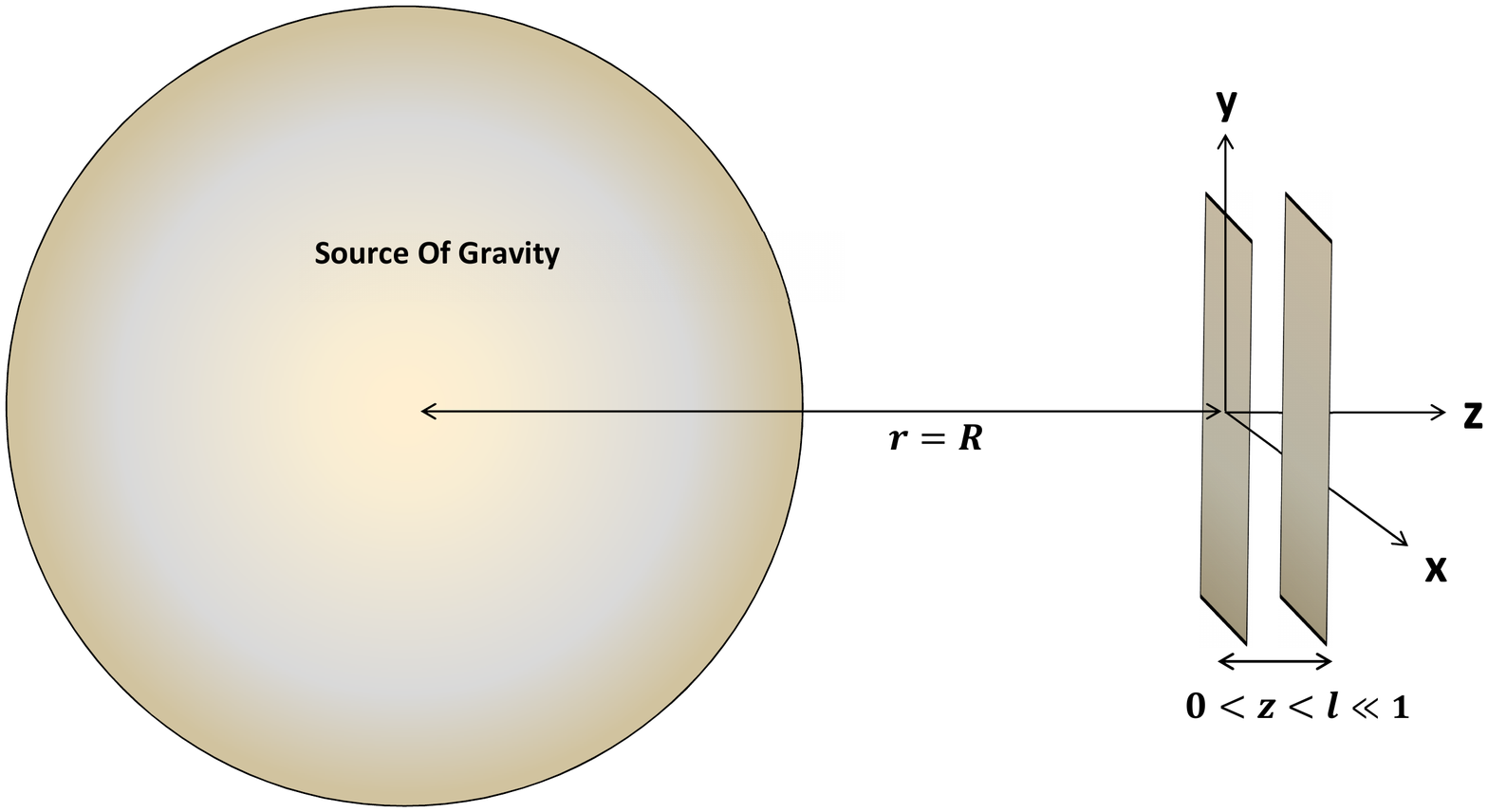}
 \caption{The Casimir apparatus far from the center of the source of a weak gravitational field.
          Z axis coincides the r axis in the equatorial plane and the plates are separated with a small coordinate distance $l<<1$. The boundaries are at z=0,z=l}
 \label{fig:apparatus}
\end{figure}

Fig. \ref{fig:apparatus} demonstrates the apparatus in a weak gravitational field. Two ideal conducting plates are in a small coordinate distance $l$ from each other. Inside the plates, the metric has a small variation relative to the flat spacetime metric. To find this variation, we adapt a rectangular coordinate system having the origin at one of the plates (the one which is closer to the source and have coordinate distance $R$ from it) and expand the metric (\ref{eq04}) inside the apparatus in the neighborhood of the point $r=R$ . The overall size of the apparatus is so small that we can assume $r=R+z$:
\begin{eqnarray}\label{eq05}
g_{\mu\nu}(r,\theta)=\eta_{\mu\nu}+h_{\mu\nu}(R+z,\theta)=\eta_{\mu\nu}+h_{\mu\nu}(R,\theta)+\frac{dh_{\mu\nu}(r,\theta)}{dr}|_{_{z=0}} z+ O(\epsilon z^2),
\end{eqnarray}
in which $R>>z, \;\ h_{\mu\nu}<<1$.  For the case of a static spacetime, the components of the metric (\ref{eq05}) can be written in the following form:
\begin{eqnarray}\label{eq06}
g_{\mu\nu}=1 +2\gamma +2 \lambda z+ O(\gamma z^2),
\end{eqnarray}
where $\gamma<1,\lambda<1$ are constants and use is made of $\Phi=\gamma+\lambda z+O(\gamma^2) \;\;\ , \gamma=\frac{-Gm}{R},\; \lambda=\frac{Gm}{R^2}$. Concerning the form of $A^i$, the above expansion satisfies $\frac{d h_{\mu\nu}(R,\theta)}{dr}|_{z=0}<1$ provided that $\theta=\pi/2$ i.e. in the equatorial plane. So (\ref{eq05}),(\ref{eq06}) are also valid for the case of the far field limit of the Kerr spacetime and we will back to it in the examples in the section VI.

Motivated by the above discussion, in this paper we analyse the general case of the spacetime of the form
\begin{eqnarray}\label{eq07}
ds^2 = (1 +2\gamma_0 +2 \lambda_0 z) dt^2 - (1+2\gamma_1 +2 \lambda_1 z)[dx^2+dy^2+dz^2],
\end{eqnarray}
in which $\gamma_0, \lambda_0, \gamma_1, \lambda_1 <1$.

To solve the Klein-Gordon equation, it is better to recast the metric (\ref{eq07}) to the known Fermi metric. We use the linearized weak field regime of general relativity and change the variables with the aid of the following gauge transformation:
\begin{eqnarray}\label{eq08}
\begin{split}
g_{\mu\nu} &=\eta_{\mu\nu}+h_{\mu\nu}, \;\; |h_{\mu\nu}|<<1, \;\;\ h^{'}_{\mu\nu} =h_{\mu\nu}-\zeta_{\mu , \nu}-\zeta_{\nu , \mu},   \\
x^{'\mu} &= x^{\mu}+\zeta^{\mu}, \\
\zeta_{t} &=(\gamma_0 + \lambda_0(z-z^{'}) )t  , \; \zeta^{t} =(\gamma_0 + \lambda_0(z-z^{'}) )t ,\\
\zeta_{x} &=-(\gamma_1 + \lambda_1 z)x  , \; \zeta^{x} =(\gamma_1 + \lambda_1 z)x ,\\
\zeta_{y} &=-(\gamma_1 + \lambda_1 z)y, \; \zeta^{y} =(\gamma_1 + \lambda_1 z)y  ,\\
\zeta_{z} &=-\gamma_1 z- \lambda_1 z^2 , \; \zeta^{z} =\gamma_1 z+ \frac{1}{2}\lambda_1 z^2.\\
\end{split}
\end{eqnarray}
which we have assumed $h^{'}_{ij}=0$ to force the spatial sector of the metric (\ref{eq07}) to be flat. More explicitly we have:
\begin{eqnarray}\label{eq09}
\begin{split}
t^{'} &=t+\gamma_0 t, \\
x^{'} &=x+(\gamma_1 + \lambda_1 z)x,  \\
y^{'} &=y+(\gamma_1 + \lambda_1 z)y , \\
z^{'} &=z+\gamma_1 z+ \frac{1}{2}\lambda_1 z^2.
\end{split}
\end{eqnarray}
the metric then takes the following form up to first order in the parameters $\gamma_0 , \lambda_0,\gamma_1 , \lambda_1$
\begin{eqnarray}\label{eq10}
ds^2 = (1 +2\lambda_0 z^{'})dt^2 - dx^{'2}-dy^{'2}-dz^{'2}.
\end{eqnarray}
The gauge transformation, however, changes our primary problem as follows. In fact according to the last equation in  (\ref{eq08}) the boundaries must be transformed from $z=0$ and $z=l$ in the spacetime (\ref{eq07}) to $z^{'}=0$ and $z^{'}=l+\gamma_1 l+ \frac{1}{2}\lambda_1 l^2$ in the spacetime (\ref{eq10}). Another change that the gauge transformation brings into the problem is the rescaling of time in (\ref{eq09}) by the factor $1+\gamma_0$. This corresponds, in turn, to dividing the mode frequencies $\omega$ by $1+\gamma_0$ because of the presence of the factor $e^{-i\omega t}$ in our solution of the Klein-Gordon equation in the next section. So Time must be re-inverted after solving the Klein-Gordon equation when we obtain the mode frequencies. \emph{The net effect of the rescaling of time is that the final mode frequencies must be multiplied by the factor $1+\gamma_0$}.
We drop the dashes $\underline{'}$ on $x^{'},y^{'},z^{'}$ from now on using the new boundary conditions instead.

\section{Exact solution to the massless Klein-Gordon equation in the Fermi metric}

The massless Klein-Gordon equation is
\begin{eqnarray}\label{eq11}
\partial_\mu[\sqrt{-\mathfrak{g}} g^{\mu\nu} \partial_\nu \Phi(x^c)]= 0
\;\;\;\;\;\;,\;\;\; \mathfrak{g}\equiv {\rm det}g^{\mu\nu}.
\end{eqnarray}
Since the spacetime is spatially flat we assume the following form for the solution
\begin{eqnarray}\label{eq12}
\Phi(x) = C e^{-i\omega t} e^{ik_x x}e^{ik_y y}Z(z),
\end{eqnarray}
where $C$ is normalization constant determined through the commutation relations
\begin{eqnarray}\label{eq13}
(\Phi_i(x) , \Phi_j(x)) = \delta_{ij}\delta({\bf k}_i - {\bf k}_j).
\end{eqnarray}
The scalar product is defined as
\begin{eqnarray}\label{eq14}
(\Phi_1 , \Phi_2) =
-i\int_{\Sigma}\Phi_1(x){\overleftrightarrow{\partial}}_{\mu}
\Phi_2^*(x)[-\mathfrak{g}_{\Sigma}(x)]^{\frac{1}{2}}n^{\mu} d\Sigma.
\end{eqnarray}
in which $n_\mu=\partial_\mu z$ and $d\Sigma$ spans the space between the plates. Under the above assumptions equation (\ref{eq11}) reads
\begin{eqnarray}\label{eq15}
(1+2\lambda z)Z''(z)+\lambda Z'(z)+(\omega^2-(1+2\lambda z)k_{\perp}^2)Z(z)=0,
\end{eqnarray}
where $'$ denotes derivation with respect to z and $k_\bot^2=k_x^2+k_y^2$. Another variable change $V(z)=\frac{Z(z)}{\sqrt{1+2\lambda z}}$ yields
\begin{eqnarray}\label{eq16}
(1+2\lambda z)V''(z)+3\lambda V'(z)+(\omega^2-(1+2\lambda z)k_{\perp}^2)V(z)=0.
\end{eqnarray}
Appearance of factor 3 in front of the second term, introduces a significant simplification when we change the variable to $T(z)=exp{(k_{\perp}z)}V(z)$. It recasts (\ref{eq16}) into
\begin{eqnarray}\label{eq17}
(1+2\lambda z)T''(z)+(3\lambda-2 k_{\perp}(1+2\lambda z))  T'(z)+(\omega^2-3\lambda k_{\perp})T(z)=0
\end{eqnarray}
A simple reparametrization of this last equation via $u=\frac{k_{\perp}}{\lambda}(1+2\lambda z)$ end up with
\begin{eqnarray}\label{eq18}
uT''(u)+(\frac{3}{2}-u)T'(u)+(\frac{\omega^2}{4 k_{\perp}\lambda}-\frac{3}{4})T(u)=0
\end{eqnarray}
This is the known Kummer's differential equation
\begin{eqnarray}\label{eq19}
uT''(u)+(B-u)T'(u)-AT(u)=0
\end{eqnarray}
in which $A=\frac{3}{4}-\frac{\omega^2}{4 k_{\perp}\lambda}\;\ , \;\ B=\frac{3}{2}$.
Kummer's differential equation is not suitable for next considerations and we transform it via $T(u)=u^{-\frac{B}{2}} e^{\frac{u}{2}} W(u)$ to another known form called Wittaker's differential equation:
\begin{eqnarray}\label{eq20}
W''(u)+(-\frac{1}{4}+\frac{B-2A}{2u}+\frac{\frac{1}{4}-\mu^2}{u^2})W(u)=0
\end{eqnarray}
in which in our case $\mu=\frac{B-1}{2}=\frac{1}{4}  \;\;\;\;\;\   , \kappa \equiv\frac{B-2A}{2}=\frac{\omega^2}{4 k_{\perp}\lambda}$.
Equation (\ref{eq20}) has two independent set of solutions $M_{\kappa,\mu}$, $W_{\kappa,\mu}$ and their asymptotic behaviour is as follows \cite{NIST}

\begin{eqnarray}\label{eq21}
M_{\kappa,\mu}(u)=\left\{
\begin{array}{cc}
\Gamma(1+2\mu)e^{\frac{u}{2}} u^{-\kappa}  \;\;\;\;\;\ u\rightarrow\infty \\
u^{\frac{3}{4}}     \;\;\;\;\;\;\;\;\;\;\;\;\;\;\;\;\;\;\;\;\;\;\;\;\    u\rightarrow 0,
\end{array}
\right.
\end{eqnarray}
\begin{eqnarray}\label{eq22}
W_{\kappa,\mu}(u)=\left\{
\begin{array}{cc}
e^{-\frac{u}{2}} u^{\kappa} \;\;\;\;\;\;\;\;\;\;\;\;\;\;\;\;\;\;\;\;\  u\rightarrow\infty \\
\frac{\Gamma(2\mu)}{\Gamma(\frac{1}{2}+\mu-\kappa)} u^{\frac{1}{2}-\mu} \;\;\;\;\;\;\;\;\ u\rightarrow 0
\end{array}
\right.
\end{eqnarray}
In general, $W_{\kappa,\mu}$ is the acceptable physical solution which is finite at infinity. In the problem under consideration we must choose a linear combination of the two, due to the fact that both are finite in between the plates. The exact mode functions are:
\begin{eqnarray}\label{eq23}
\phi_{\kappa}(u)=u^{-\frac{1}{4}} [A(\omega,k_{\bot}) W_{\kappa,\frac{1}{4}}(u)+B(\omega,k_{\bot}) M_{\kappa,\frac{1}{4}}(u)] e^{-i\omega t-ik_xx -ik_yy}\;\;\;\;\;\;\;\;\
\end{eqnarray}
The asymptotic form of (\ref{eq23}) for small value of $\lambda_0$ can be written as follows (see Appendix A)
\begin{eqnarray}\label{eq24}
\phi_{\kappa}(z)=C_0(\omega,k_\perp) (g_{00}S(z))^{-\frac{1}{4}} sin(\int_0^z \sqrt{S}dz+\phi_0)  e^{-i\omega t-ik_{x} x -ik_{y} x}
\end{eqnarray}
in which $S=\frac{\omega^2}{g_{00}}-k_\perp^2, \;\ g_{00}=1+2\lambda_0 z$. In the next section we use this asymptotic form to extract mode frequencies for both Neumann and Dirichlet boundary conditions on the plates.

\section{Mode frequencies for Neumann and Dirichlet boundary conditions for the scalar field}
We put the approximation $\int \sqrt{S}dz \simeq \sqrt{b}z+\frac{a}{4\sqrt{b}}z^2 \;\;\ , a=-2\omega^2\lambda_0 \;\ ,b=\omega^2-k_\perp^2$ into (\ref{eq24}). From the Dirichlet boundary condition $\phi_\kappa(z=0)=0$ we have $\phi_0=0$ and from $\phi_\kappa(z=l+\gamma_1 l+ \frac{1}{2}\lambda_1 l^2)=0$ we have:
\begin{eqnarray}\label{eq25}
\int_0^{l+\gamma_1 l+ \frac{1}{2}\lambda_1 l^2} \sqrt{S}dz=n\pi.
\end{eqnarray}
After careful expansion of (\ref{eq25}), the mode frequencies proved to satisfy the following relation:
\begin{eqnarray}\label{eq26}
\omega^2 \{1-\lambda_0[l+\gamma_1 l+ \frac{1}{2} \lambda_1 l^2]\}=\sqrt{k_\perp^2+(\frac{n\pi}{l+\gamma_1 l+ \frac{1}{2} \lambda_1 l^2})^2}
\end{eqnarray}
Note that the factor $l+\gamma_1 l+ \frac{1}{2}\lambda_1 l^2=l_P=\int_0^l \sqrt{g_{33}}dz=\int_0^l \sqrt{1+2\gamma_1 + 2\lambda_1 z}dz$ is nothing but the \emph{proper distance} between the plates and so we have
\begin{eqnarray}\label{eq27}
\omega=\omega_0 (1+\lambda_0\frac{l_p}{2} )
\end{eqnarray}
in which $\omega_0=\sqrt{k_\perp^2+(\frac{n\pi}{l_P})^2} \;\ , \;\ n=0,1,2,...$ denotes proper (or the corresponding flat space ) mode frequencies in the local Lorentz frame of an observer comoving with the plates. As stated in the previous section, the final mode frequencies will be obtained by multiplication of the factor $1+\gamma_0$ due to rescaling of time during the gauge transformation (\ref{eq09}). So we have the final mode frequencies inside the Casimir apparatus for the spacetime (\ref{eq07}):
\begin{eqnarray}\label{eq28}
\omega=\omega_0 (1+\gamma_0+\lambda_0\frac{l_p}{2})
\end{eqnarray}
Mode frequencies are influenced only by $g_{00}$ component of the metric from the point of view of a proper observer.

The Neumann boundary condition $\partial_z\phi|_{z=0}=0$ imposed on (\ref{eq24}) gives $\phi_0=\frac{\pi}{2}$:
\begin{eqnarray}\label{eq29}
\frac{d\phi}{dz}|_{z=0}=0\;\;\ \Rightarrow tan(\int_0^0 \sqrt{S}dz+\phi_0)|_{z=0}=\frac{4\sqrt{S}(g_{00}S)}{\frac{d}{dz}(g_{00}S)}|_{z=0}=\frac{4b}{O(\lambda_0)}\rightarrow \infty
\end{eqnarray}
Another Neuman boundary condition $\partial_z\phi|_{z=l}=0$ end up with
\begin{eqnarray}\label{eq30}
\frac{d\phi}{dz}|_{z=l}=0 \;\;\ \Rightarrow cot(\int_0^l \sqrt{S}dz)|_{z=l}=\frac{4\sqrt{S}(g_{00}S)}{\frac{d}{dz}(g_{00}S)}|_{z=l}=\frac{4b}{O(\lambda_0)}\rightarrow \infty
\end{eqnarray}
which in turn result in (\ref{eq26}) and (\ref{eq27}) again.

\section{Generalization of the formalism when the electromagnetic field is present inside the plates}
The electromagnetic field has two physical degrees of freedom and it is known that in the Rindler spacetime the electromagnetic field can be represented in terms of two scalar fields satisfying the Klein-Gordon equation separately \cite{Deutch}. The photon propagator and the energy-momentum tensor of the electromagnetic field in a weak gravitational of the Fermi spacetime has also been obtained in \cite{Bimonte}. However in \cite{Bimonte}, the computations has been done through a lengthy and cumbersome method of the green functions. In a paper by the author \cite{NouriNazari2}, it is shown that the energy density (i.e. the 0-0 component of the energy-momentum tensor) of the electromagnetic field in Fermi spacetime is exactly the some of the energy density of the two scalar fields mentioned in \cite{NouriNazari2}. The method was done without any address to the Green function method frequently used in the literature. Here, we briefly review the relationship between the two scalar fields and the electromagnetic field in Fermi spacetime.

The spin one vector field in curved spacetime in the Lorentz gauge satisfies
\begin{eqnarray}\label{eq31}
\square A^{\mu} + R^\mu_\nu A^\nu=0 \;\;\;\   , \;\;\;
\nabla_\mu A^{\mu}=0 \;\;\;\   , \;\;\;\mu = 0,1,2,3.
\end{eqnarray}
It can be shown that in Fermi metric, the Ricci tensor satisfies $R_{\mu\nu}=O(\lambda^2)$ and the second term of the wave equation (\ref{eq31}) must be ignored. Furthermore, because the metric (\ref{eq10}) is spatially flat, the Lorentz gauge in (\ref{eq31}) can be broken into two independent parts \cite{Deutch}:
\begin{eqnarray}\label{eq32}
\begin{split}
\nabla^a A_a=0 \;,\;a=0,3\equiv(t,z), \\
\nabla^i A_i=0 \;,\;i=1,2\equiv(x,y).
\end{split}
\end{eqnarray}
In which
\begin{eqnarray}\label{eq33}
A_i=\varepsilon_{ij}\nabla^j\phi \;\   , \;\;\;
A_a=\varepsilon_{ab}\nabla^b\psi.
\end{eqnarray}
and
\begin{eqnarray}\label{eq34}
\epsilon_{ij}=\left(
                \begin{array}{cc}
                   0 & 1 \\
                   -1 & 0 \\
                \end{array}
              \right) \;\; ,\;\;  \epsilon_{ab}=\left(
                                                   \begin{array}{cc}
                                                       0 & 1+\lambda z \\
                                                       -1-\lambda z & 0 \\
                                                   \end{array}
                                                \right).
\end{eqnarray}
We know also that both $\psi$ and $\phi$ satisfy the Klein-Gordon eqaution separately(see the appendix in \cite{NouriNazari2}). Boundary condition for the electric field on the plates is $\textbf{E}_\perp(z=0)=E_z(z=0)=0$ and $\textbf{E}_\perp(z=l)=E_z(z=l)=0$ which in turn can be recast into boundary conditions on $\psi$ and $\phi$ according to equation (\ref{eq33}). In \cite{NouriNazari2} it has been shown that boundary conditions for the electric and magnetic fields return a Dirichlet boundary condition on $\phi$ and a Neumann boundary condition on $\psi$. We have shown in previous section that both of this conditions will be ended up to a same frequency shift. As a result mode frequencies in (\ref{eq28}) are also valid for the electromagnetic field.

\section{The energy-momentum tensor}
This section has three subsections. In the first subsection, the electromagnetic energy-momentum tensor will be represented in terms of the energy momentum of the scalar fields mentioned in (\ref{eq33}). In the other two sections the energy momentum tensor of the scalar and vector fields will be calculated.
 \subsection{The relationship between the energy-momentum tensor of the scalar and vector fields}
The vacuum expectation value of the quantum energy-momentum tensor is defined as
\begin{eqnarray}\label{eq35}
<0|T_{\mu\nu}|0>=\sum_{\mathbf{k}}T_{\mu\nu}[\phi_{\mathbf{k}},\phi^*_{\mathbf{k}}].
\end{eqnarray}
The classical energy-momentum tensors for the scalar and vector fields are
\begin{eqnarray}\label{eq36}
\begin{split}
T^{Scalar}_{\mu\nu} &=\partial_\mu\phi\partial_\nu\phi-\frac{1}{2}g_{\mu\nu}g^{\lambda\theta}\partial_\lambda\phi\partial_\theta\phi,\\
T^{vector}_{\mu\nu} &=T_{\mu\nu}^{Ghost}+T_{\mu\nu}^{Gauge}+T_{\mu\nu}^{Maxwell} ,  \\
T_{\mu\nu}^{Ghost} &= \zeta^{-1}[ A_\mu A^\rho_{;\rho\nu}- A^\rho_{;\rho\mu}A_\nu-g_{\mu\nu}\{A^\rho A^\theta_{;\theta\rho}+\frac{1}{2}(A^\rho_{;\rho})^2\}] ,\\
T_{\mu\nu}^{Gauge} &=-c^*_{;\mu}c_{;\nu}-c^*_{;\nu}c_{;\mu}-g_{\mu\nu}g^{\lambda\theta}c_{;\lambda}c_{;\theta} , \\
T_{\mu\nu}^{Maxwell} &=\frac{1}{4}g_{\mu\nu}F^{\lambda\theta}F_{\lambda\theta}-F_{\mu^\theta}F_{\theta\nu}.
\end{split}
\end{eqnarray}

As is well-known, in the quantum level, the contributions of ghost and gauge fields in the electromagnetic energy-momentum tensor cancel each other \cite{Bimonte} and we only concern the Maxwell sector of the energy-momentum tensor. Expansion of $T_{\mu\nu}^{Maxwell}$ in terms of scalar fields $\psi$ and $\phi$ shows that (see Appendix B)
\begin{eqnarray}\label{eq37}
\begin{split}
T_{00}^{Maxwell} &=-\frac{1}{2}(E^2+(1+2\lambda_0z)B^2),  \\
T_{0i}^{Maxwell} &=-(\overrightarrow{E}\times \overrightarrow{B})_i ,\\
T_{ij}^{Maxwell} &=\frac{1}{2}(1-2\lambda_0z)(E^2+(1+2\lambda_0z)B^2)g_{ij}-(1-2\lambda_0z)E_iE_j-B_iB_j,
\end{split}
\end{eqnarray}
in which $E^2=g_{ij}E^iE^j, \;\;\ B^2=g_{ij}B^iB^j, \;\;\ g_{ij}=-\delta_{ij}, \;\;\  E_i=F_{0i}, \;\;\ F_{ij}=\varepsilon_{ijk}B^k $ . Using $F_{\mu\nu}=A_{\mu,\nu}-A_{\nu,\mu}$ and (\ref{eq32}) we have:
\begin{eqnarray}\label{eq38}
\begin{split}
F_{\tau x}&=-ik_x\sqrt{1+2\lambda z}\partial_z\psi+\omega k_y\phi,\\
F_{\tau y} &=-ik_y\sqrt{1+2\lambda z}\partial_z\psi-\omega k_x\phi, \\
F_{\tau z} &=-\sqrt{1+2\lambda z}k_\perp^2 \psi,  \\
F_{yz} &=ik_x\partial_z\phi+\omega (1-2\lambda z)\sqrt{1+2\lambda z} k_y\psi, \\
F_{xz}&=-ik_y\partial_z\phi+\omega (1+2\lambda z)\sqrt{1+2\lambda z} k_x\psi, \\
F_{xy} &=k_\perp^2 \phi, \\
\overrightarrow{E} &=(F_{\tau\ x},F_{\tau\ y},F_{\tau\ z}), \\
\overrightarrow{B} &=(-F_{yz},F_{xz},-F_{xy})
\end{split}
\end{eqnarray}
in which we have used the general form of the wave function (\ref{eq12}). Quadratic products of fields $E^2 \;\ ,B^2\;\ ,E_iE_j $ and $B_iB_j$ produce terms like $\psi\partial_z\phi^*, \;\ \phi\partial_z\psi^*,\;\ \phi\psi^*$ which have no contribution when the expectation value is taken because of the fact that $\psi$ and $\phi$ are not correlated and belongs to independent Hilbert spaces. We calculate $0-0$ component of the energy-momentum tensor for the electromagnetic field:
\begin{eqnarray}\label{eq39}
<0|T_{00}^{\phi}|0> &=\sum_\omega\int d^2k_\perp\{\frac{1}{2}(\omega^2+(1+2\lambda z)k_\perp^2)|\phi|^2+\frac{1}{2}(1+2\lambda z)|\partial_z\phi|^2\} \end{eqnarray}
\begin{eqnarray}\label{eq40}
\begin{split}
<0|T_{00}^{Max.}|0> &=-\sum_\omega\int d^2k_\perp\{\frac{1}{2}(<0|E^2|0>+(1+2\lambda z)<0|B^2|0>)\}  \\
 &=\sum_\omega\int d^2k_\perp\{ \frac{1}{2}(<0|F_{\tau x}^2+F_{\tau  y}^2+F_{\tau  z}^2|0>+(1+2\lambda z)<0|F_{yz}^2+F_{xz}^2+F_{xy}^2|0>) \} \\
 &=\sum_\omega\int d^2k_\perp \{k_\perp^2[\frac{1}{2}\{(\omega^2+(1+2\lambda z)k_\perp^2)|\phi|^2+(1+2\lambda z)|\partial_z\phi|^2\} \\
 &+\frac{1}{2}\{(\omega^2+(1+2\lambda z)k_\perp^2)|\psi|^2+(1+2\lambda z)|\partial_z\psi|^2\}] \} \\
&=k_\perp^2[<0|T_{00}^{\phi}|0>+<0|T_{00}^{\psi}|0>]
\end{split}
\end{eqnarray}
After a lengthy but straightforward calculation, other components of the energy-momentum tensor in both sides are related to each other as follows:
\begin{eqnarray}\label{eq41}
\begin{split}
<0|T_{00}^{Maxwell}|0> &= k_\perp^2\{ <0|T_{00}^{\phi}|0> + <0|T_{00}^{\psi}|0> \} ,\\
<0|T_{11}^{Maxwell}|0> &= -k_\perp^2\{ <0|T_{11}^{\phi}|0> + <0|T_{11}^{\psi}|0> \} +2(1-2\lambda z)k_x^2\{<0|T_{00}^{\phi}|0>+<0|T_{00}^{\psi}|0>\},\\
<0|T_{22}^{Maxwell}|0> &= -k_\perp^2\{ <0|T_{22}^{\phi}|0> + <0|T_{22}^{\psi}|0> \} +2(1-2\lambda z)k_y^2\{<0|T_{00}^{\phi}|0>+<0|T_{00}^{\psi}|0>\}, \\
<0|T_{33}^{Maxwell}|0> &= k_\perp^2\{ <0|T_{33}^{\phi}|0> + <0|T_{33}^{\psi}|0> \}, \\
<0|T_{01}^{Maxwell}|0> &= k_\perp^2\{ <0|T_{01}^{\phi}|0> + <0|T_{01}^{\psi}|0> \}, \\
<0|T_{02}^{Maxwell}|0> &= k_\perp^2\{ <0|T_{02}^{\phi}|0> + <0|T_{02}^{\psi}|0> \}, \\
<0|T_{03}^{Maxwell}|0> &= -k_\perp^2\{ <0|T_{03}^{\phi}|0> + <0|T_{03}^{\psi}|0> \},\\
<0|T_{12}^{Maxwell}|0> &= -k_\perp^2\{ <0|T_{12}^{\phi}|0> + <0|T_{12}^{\psi}|0> \} +2(1-2\lambda z)\{<0|T_{00}^{\phi}|0>+<0|T_{00}^{\psi}|0>\}, \\
<0|T_{23}^{Maxwell}|0> &= -k_\perp^2\{ <0|T_{23}^{\phi}|0> + <0|T_{23}^{\psi}|0> \}, \\
<0|T_{13}^{Maxwell}|0> &= -k_\perp^2\{ <0|T_{13}^{\phi}|0> + <0|T_{13}^{\psi}|0> \}.
\end{split}
\end{eqnarray}
Note that $k_\perp^2$ can be absorbed into $C$.

 \subsection{Energy density for Dirichlet and Neumann scalar fields}
This section is devoted to the calculation of the energy density for the Casimir apparatus via the direct method without any reference to the traditional Green function method. Using the approximations
\begin{eqnarray}\label{eq42}
\begin{split}
(g_{00}S)^{-\frac{1}{4}} & \simeq 1-(\frac{\lambda}{2}+\frac{a}{4b})z, \\
\int \sqrt{S}dz &\simeq \sqrt{b}z+\frac{a}{4\sqrt{b}}z^2 \;\;\ , a=-2\omega^2\lambda \;\ ,b=\omega^2-k_\perp^2,  \\
sin(\sqrt{b}z+\frac{a}{4\sqrt{b}}z^2) &=sin(\sqrt{b}z)+\frac{a}{4\sqrt{b}}z^2 cos(\sqrt{b}z), \\
cos(\sqrt{b}z+\frac{a}{4\sqrt{b}}z^2) &=cos(\sqrt{b}z)-\frac{a}{4\sqrt{b}}z^2 sin(\sqrt{b}z).
\end{split}
\end{eqnarray}
We expand the wave function (\ref{eq24}) and find up to first order in $\lambda$:
\begin{eqnarray}\label{eq43}
\begin{split}
Z(z) &=Z_0 \{[1-(\frac{\lambda}{2}+\frac{a}{4b})z] sin(\sqrt{b}z)+\frac{a}{4\sqrt{b}}z^2 cos(\sqrt{b}z)\} \;\;\ , Dirichlet \\
Z(z) &=Z_0 \{[1-(\frac{\lambda}{2}+\frac{a}{4b})z] cos(\sqrt{b}z)-\frac{a}{4\sqrt{b}}z^2 sin(\sqrt{b}z)\} \;\;\ , Newmann
\end{split}
\end{eqnarray}
$Z_0$ can be absorbed also in $C_0$.

The energy density is defined by $\varepsilon=n^\mu n^\nu <0|T_{\mu\nu}^{\phi}|0>$ where $n^\mu$ is the lapse vector normal to the hypersurface $z=constant$ i.e. $n^\mu=\partial_z z=\frac{1}{1+2\lambda_0z}(0,0,0,1)$. The mean energy density so has the following form:
\begin{eqnarray}\label{eq44}
\begin{split}
\overline{\varepsilon} &=\frac{1}{l} \int_0^l n^\mu n^\nu <0|T_{\mu\nu}^{\phi}|0> dz \\
&=\sum_\omega\int d^2k_\perp \frac{1}{l} \int_0^l \{\frac{1}{2}(\omega^2+(1+2\lambda_0 z)k_\perp^2)Z(z)^2+\frac{1}{2}(1+2\lambda_0 z)|\partial_zZ(z)|^2\} \frac{dz}{(1+2\lambda_0 z)} \\
&\equiv\sum_\omega\int d^2k_\perp H(\omega,k_\perp)
\end{split}
\end{eqnarray}
Calculating the factor $H(\omega,k_\perp)$ for both wave functions in (\ref{eq45}) results in
\begin{eqnarray}\label{eq45}
\begin{split}
H^{Dirichlet}(\omega,k_\perp) = & C^2 \{ \frac{1}{2}\omega^2 F_1 +\frac{1}{2}k_\perp^2 F_2 +\frac{1}{2} F_3 \}, \\
F_1 =& \{  \frac{z}{2}-\frac{sin(2\sqrt{b}z)}{4\sqrt{b}}-(3\lambda+\frac{a}{2b})[\frac{z^2}{4}-\frac{z sin(2\sqrt{b}z)}{4\sqrt{b}}-\frac{ cos(2\sqrt{b}z)}{8b}]  \\ & +\frac{a}{4\sqrt{b}}[\frac{z}{2b}sin(2\sqrt{b}z) +(\frac{1}{2b}-z^2)\frac{cos(2\sqrt{b}z)}{2\sqrt{b}}]  \}|_\alpha^\beta, \\
F_2 =& \{\frac{z}{2}-\frac{sin(2\sqrt{b}z)}{4\sqrt{b}}-(\lambda+\frac{a}{2b})[\frac{z^2}{4}-\frac{z sin(2\sqrt{b}z)}{4\sqrt{b}}-\frac{ cos(2\sqrt{b}z)}{8b}] \\ & +\frac{a}{4\sqrt{b}}[\frac{z}{2b}sin(2\sqrt{b}z)+(\frac{1}{2b}-z^2)\frac{cos(2\sqrt{b}z)}{2\sqrt{b}}]  \}|_\alpha^\beta,  \\
F_3 =& \{ b[\frac{z}{2}+\frac{sin(2\sqrt{b}z)}{4\sqrt{b}}+(\frac{a}{2b}-\lambda)[\frac{z^2}{4}+\frac{z sin(2\sqrt{b}z)}{4\sqrt{b}}+\frac{ cos(2\sqrt{b}z)}{8b}] ] \\ & -\frac{a\sqrt{b}}{4} [\frac{z}{2b}sin(2\sqrt{b}z)+(\frac{1}{2b}-z^2)\frac{cos(2\sqrt{b}z)}{2\sqrt{b}}] +(\frac{\lambda}{4}+\frac{a}{8b}) cos(2\sqrt{b}z)\}|_\alpha^\beta. \\
\end{split}
\end{eqnarray}
\begin{eqnarray}\label{eq46}
\begin{split}
H^{Neuman}(\omega,k_\perp) = & C^2 \{ \frac{1}{2}\omega^2 F_1 +\frac{1}{2}k_\perp^2 F_2 +\frac{1}{2} F_3 \}, \\
F_1 =& \{  \frac{z}{2}+\frac{sin(2\sqrt{b}z)}{4\sqrt{b}}-(3\lambda+\frac{a}{2b})[\frac{z^2}{4}+\frac{z sin(2\sqrt{b}z)}{4\sqrt{b}}+\frac{ cos(2\sqrt{b}z)}{8b}]  \\ & -\frac{a}{4\sqrt{b}}[\frac{z}{2b}sin(2\sqrt{b}z) +(\frac{1}{2b}-z^2)\frac{cos(2\sqrt{b}z)}{2\sqrt{b}}]  \}|_\alpha^\beta, \\
F_2 =& \{\frac{z}{2}+\frac{sin(2\sqrt{b}z)}{4\sqrt{b}}-(\lambda+\frac{a}{2b})[\frac{z^2}{4}+\frac{z sin(2\sqrt{b}z)}{4\sqrt{b}}+\frac{ cos(2\sqrt{b}z)}{8b}] \\ & -\frac{a}{4\sqrt{b}}[\frac{z}{2b}sin(2\sqrt{b}z)+(\frac{1}{2b}-z^2)\frac{cos(2\sqrt{b}z)}{2\sqrt{b}}]  \}|_\alpha^\beta,  \\
F_3 =& \{ b[\frac{z}{2}-\frac{sin(2\sqrt{b}z)}{4\sqrt{b}}+(\frac{a}{2b}-\lambda)[\frac{z^2}{4}-\frac{z sin(2\sqrt{b}z)}{4\sqrt{b}}-\frac{ cos(2\sqrt{b}z)}{8b}] ] \\ & +\frac{a\sqrt{b}}{4} [\frac{z}{2b}sin(2\sqrt{b}z)+(\frac{1}{2b}-z^2)\frac{cos(2\sqrt{b}z)}{2\sqrt{b}}] -(\frac{\lambda}{4}+\frac{a}{8b}) cos(2\sqrt{b}z)\}|_\alpha^\beta.
\end{split}
\end{eqnarray}

Simplification of the terms like $cos(2\sqrt{b}z)$ and $sin(2\sqrt{b}z)$ is possible using the fact that in boundaries the wave function (\ref{eq24}) must vanish and:
\begin{eqnarray}\label{eq47}
\begin{split}
\int_0^z \sqrt{S}dz=n\pi-\Phi_0  \;\ \Rightarrow \;\ \sqrt{b}z+\frac{a}{4\sqrt{b}}z^2=n\pi -\Phi_0,
\end{split}
\end{eqnarray}
in which $z$ takes one of two boundary values $z=0$ and $z=l+\gamma_1 l+ \frac{1}{2}\lambda_1 l^2$. Equation (\ref{eq47}) result in
\begin{eqnarray}\label{eq48}
\begin{split}
sin(2\sqrt{b}z)=-\frac{a}{2\sqrt{b}}z^2-\Phi_0 \;\ , \;\ cos(2\sqrt{b}z)\simeq1, \\
\end{split}
\end{eqnarray}
and the final result is as follows
\begin{eqnarray}\label{eq49}
H^{Neuman}(\omega,k_\perp) = C^2 \frac{\omega^2}{2}\{ l-[\lambda +\frac{a}{4b}](\beta^2-\alpha^2) \}.
\end{eqnarray}
Exactly the same $H$ is obtained for Dirichlet boundary condition. The constant $C$ was defined in (\ref{eq07}) and can be determined simply as:
\begin{eqnarray}\label{eq50}
\begin{split}
C^2 &=\frac{1}{2(2\pi)^2\omega} \{\int_0^l \{[1-(2\lambda+\frac{a}{2b})z] sin^2(\sqrt{b}z)+\frac{a}{4\sqrt{b}}z^2 sin(2\sqrt{b}z)\}dz\}^{-1} \;\ , Dirichlet \\
C^2 &=\frac{1}{2(2\pi)^2\omega} \{\int_0^l \{[1-(2\lambda+\frac{a}{2b})z] cos^2(\sqrt{b}z)-\frac{a}{4\sqrt{b}}z^2 sin(2\sqrt{b}z)\}dz\}^{-1} \;\ , Neumann
\end{split}
\end{eqnarray}
Just like $H$, the constant $C$ has also the same form for both boundary conditions up to first order in $\lambda$
\begin{eqnarray}\label{eq51}
C^2 =\frac{1}{2(2\pi)^2\omega} \{ \frac{l}{2}-(2\lambda+\frac{a}{2b})\frac{(\beta^2-\alpha^2)}{4} \}^{-1}
\end{eqnarray}
The final result for the energy-density after all is:
\begin{eqnarray}\label{eq52}
\begin{split}
\overline{\varepsilon} &=\sum_\omega\int d^2k_\perp\frac{\omega}{2(2\pi)^2} \\
&=\sum_\omega\int d^2k_\perp\frac{\omega_0 (1+\gamma_0+\lambda_0\frac{l_P}{2})}{2(2\pi)^2} \\
&=(1+\gamma_0+\lambda_0\frac{l_P}{2})\sum_\omega\int d^2k_\perp\frac{\omega_0 }{2(2\pi)^2}.
\end{split}
\end{eqnarray}
in which $\overline{\varepsilon}_0=\sum_\omega\int d^2k_\perp\frac{\omega_0 }{2(2\pi)^2}$ is the corresponding flat spacetime Casimir energy density $\overline{\varepsilon}_0=-\frac{\pi^2}{1440 l^4}$ \cite{Birrell}.

Up to now, we have shown that for Neumann and Dirichlet boundary conditions there exists the relation $<0|T^\psi_{00,Fermi}|0>=(1+\gamma_0+\lambda_0\frac{l_P}{2})<0|T^\psi_{00,Flat}|0>$ between flat and curved energy density contents of the Casimir apparatus in the weak spacetime of the metric (\ref{eq07}).

 \subsection{Energy density for the electromagnetic field}
Using the first equation of (\ref{eq41}) it is evident that the same shift in the energy density obtained in (\ref{eq52}) holds also for the electromagnetic field :
\begin{eqnarray}\label{eq53}
\begin{split}
<0|T_{00}^{Maxwell}|0> &= <0|T_{00}^{\phi}|0> + <0|T_{00}^{\psi}|0> \\
&= \{<0|T_{00,flat}^{\phi}|0> + <0|T_{00,flat}^{\psi}|0>\}(1+\gamma_0+\lambda_0\frac{l_P}{2}) \\
&=\{-\frac{\pi^2}{1440 l^4}-\frac{\pi^2}{1440 l^4}\}(1+\gamma_0+\lambda_0\frac{l_P}{2}) \\
&=-\frac{\pi^2}{720 l^4}(1+\gamma_0+\lambda_0\frac{l_P}{2})
\end{split}
\end{eqnarray}
In the next section we will analyse some well-known weak gravitational fields and find the parameters $\gamma_0,\gamma_1,\lambda_0,\lambda_1$ in each case.
 \subsection{Notes on the divergences}
The problem of the divergences near a perfect generic conductor first studied systematically by Deutsch and Candelas \cite{Deutsch}. They found that the energy-momentum tensor near the surface behaves like:
\begin{eqnarray}\label{eq54}
<0|T_{00}|0> = \frac{c_1}{\epsilon^4} +\frac{c_2}{\epsilon^3}+ ... ,
\end{eqnarray}
where $c_1,c_2$ are constants and $\epsilon$ is the distance from the surface of the ideal boundary. In case where there is a conformal invariance in the action $c_1=0$. The divergences originates from the unphysical nature of classical ideal conductor boundary conditions. It has been shown \cite{Deutsch} that we can remove the infinities of the total energy (and not the energy-momentum tensor) of the plates using the zeta function regularization unless for the case that the zeta function has poles itself. On the other hand, the cut off regularization method suggests the removal of the divergences via \emph{ad hoc} although in this method it still remains a logarithmic ambiguity \cite{Estrada} in the energy density. For the imperfect conductors (more realistic boundary conditions) we can easily remove the divergences introducing some suitable cut off frequencies although the boundary effect may become quit large (but finite)\cite{Milton1,Deutsch}.

Now the question is that what happens when we go to curved spacetime?. Does the surface divergences of the energy-momentum tensor are ignorable in the semi-classical Einstein's equations?. The answer is negative according to \cite{Deutsch}. In \cite{Estrada} however, the authors has found a way to get ride of the surface divergences of the Einstein's field equations for the case the boundary is a parallelepiped. They have used a suitable cut off along with the so-called Estrada-Kanwal distribution theory of asymptotics to regularize/renormalize the infinities and show that the energy-momentum tensor near a plane boundary, as a source, converges to a consistent theory when the cutoff is removed.
Remarkably, the process of curing the divergences in curved spacetime has been continued by Milton et al \cite{Milton1} where they have shown that most of the energy between the plates is restored near the surfaces and a part of it resides exactly on the plates. They finally have shown that this energies responds to gravity just like any other finite energy following the newtonian relation $F=ma=-Mg$. This is expectable as this large energies are simply a part of the total energy of the system.

We show here that, in our case, the divergences do not present in the first order of approximation that we have used here and they appear only in the higher orders of approximation. To do so, we write down the explicit structure of possible divergences in the Einstein's field equations along the lines depicted in \cite{Birrell}. The one-loop effective action $W$ for the semi-classical theory of gravity is:
\begin{eqnarray}\label{eq55}
\begin{split}
&R_{\mu\nu}-\frac{1}{2}Rg_{\mu\nu}+\Lambda g_{\mu\nu}=-\frac{8\pi G}{c^4}<|T_{\mu\nu}|>, \\
&W=\int \sqrt{-g}L_{eff.}(x)d^nx, \\
&L_{eff.}(x)=\frac{i}{2} \lim_{x \to x^{'}} \int_{m^2}^\infty dm^2G_F^{D.S.}(x,x^{'}),
\end{split}
\end{eqnarray}
in which $G_F^{D.S.}(x,x^{'})$ is the DeWitt-Schwinger-Feynmann's propagator. Using the DeWitt-Schwinger representation of the action, the asymptotic expansion of $L_{eff.}$ is as follows:
\begin{eqnarray}\label{eq56}
L_{eff.}(x)=\frac{1}{2} (4\pi)^{-\frac{n}{2}} (\frac{m}{\mu})^{n-4} \sum_{j=0}^\infty a_j(x) m^{4-2j} \Gamma(j-\frac{n}{2}), 
\end{eqnarray}
in which $\mu$ is a length scale to fix the dimensional issues. The potentially divergent part of the effective action (the first three terms) \cite{Birrell} reads :
\begin{eqnarray}\label{eq57}
\begin{split}
L_{div}=- (4\pi)^{\frac{n}{2}} \{\frac{1}{n-4}+\frac{1}{2}[\gamma+\ln{\frac{m^2}{\mu^2}}] \} [\frac{1}{n(n-2)}4m^2a_0-\frac{1}{(n-2)}2m^2a_1+a_2] \\.
\end{split}
\end{eqnarray}
in which $m$ is the mass of the scalar field(m=0 in our case). The coefficients $a_0,a_1,a_2$ are:
\begin{eqnarray}\label{eq58}
\begin{split}
a_0(x)=1, \; a_1(x)=\frac{1}{6}R, \;  a_2(x)=\frac{1}{180}R_{\alpha\beta\gamma\delta}R^{\alpha\beta\gamma\delta}-\frac{1}{180}R_{\alpha\beta}R^{\alpha\beta}-\frac{1}{6}\Box R+\frac{1}{72}R^2, \\
\end{split}
\end{eqnarray}
in which $R_{\alpha\beta\gamma\delta}$ is the Riemann's curvature tensor. The total gravitational lagrangian density can be shown to have the form (see (6.49) in \cite{Birrell}):
\begin{eqnarray}\label{eq59}
\begin{split}
\mathscr{L}=-(A+\frac{\Lambda_B}{8\pi G_B}) &+(B+\frac{1}{16\pi G_B})R-\frac{1}{(4\pi)^{n/2}}\{\frac{1}{n-4}+\frac{1}{2}[\gamma+\ln{\frac{m^2}{\mu^2}}] \} a_2, \\
&A=\frac{4m^4}{(4\pi)^{\frac{n}{2}} n(n-2)} \{\frac{1}{n-4}+\frac{1}{2}[\gamma+\ln{\frac{m^2}{\mu^2}}] \} \\
&B=\frac{m^2}{(4\pi)^{3\frac{n}{2}} (n-2)} \{\frac{1}{n-4}+\frac{1}{2}[\gamma+\ln{\frac{m^2}{\mu^2}}] \}. \\
\end{split}
\end{eqnarray}
However, in the massless case of ours, the only non-vanishing potentially ultraviolent term is the one related to $a_2$ in (\ref{eq52}) and $A,B$ in (\ref{eq52}) vanish (see (6.101) in \cite{Birrell}). Our calculation for the metric (\ref{eq07}) shows that $a_2$ is of second order of approximation:
\begin{eqnarray}\label{eq60}
\begin{split}
&R_{1212}=R_{1313}=-\frac{\lambda_0\lambda_1}{4(1+2\gamma_1 +2 \lambda_1 z)}=O(\lambda_0^2)   ,\\ &R_{1414}=\frac{\lambda_0^2}{4(1+2\gamma_0 +2 \lambda_0 z)}+\frac{\lambda_0\lambda_1}{4(1+2\gamma_1 +2 \lambda_1 z)}=O(\lambda_0^2), \\
&-2R_{2323}=R_{2424}=R_{3434}=-\frac{\lambda_1^2}{2(1+2\gamma_1 +2 \lambda_1 z)}=O(\lambda_1^2), \\
& R_{11}=R_{22}=R_{33}\simeq O(\lambda_0^2),\; R_{44}\simeq O(\lambda_0^2),\; R\simeq O(\lambda_0^2),\; \Box R\simeq O(\lambda_0^2).
\end{split}
\end{eqnarray} 
So the total lagrangian density (\ref{eq59}) reduces to the the standard bare density $\frac{1}{16\pi G_B}R$. In conclusion, the potentially divergent term $a_2$ vanishes within the first order of approximation.
\section{Examples:finding coefficients $\gamma_0,\gamma_1,\lambda_0,\lambda_1$}
In this section number of spacetimes are investigated and the parameters appeared in (\ref{eq28}),(\ref{eq52}),(\ref{eq53}) will be found. To this end, we will try to find their weak field form according to (\ref{eq03}),(\ref{eq04}) and (\ref{eq07}).

\subsection{Electromagnetic Casimir energy density for the far field limit of the Kerr spacetime}
Recently Bezerra et al \cite{Bezerra} studied the Apparatus for the scalar fields in the weak field limit of the Kerr spacetime in the equatorial plane. In that work the apparatus co-rotates with the local angular velocity of the spacetime i.e. the measurements had been assumed to be done in the point of view of a zero angular momentum observer(ZAMO). They found the metric inside the apparatus through some two stage successive approximation method as follows:
\begin{eqnarray}\label{eq60}
\begin{split}
ds^2 \approx & (1 +2b\Phi_0)dt^2 - (1 -2\Phi_0)[dx^2+dy^2+dz^2]\;\;\; 0-th\; order\; of\; approximation ,\\
ds^2 \approx & (1 +2b\lambda z)dt^2 - (1 -2\lambda z)[dx^2+dy^2+dz^2]\;\;\; first\;order\;of\;approximation.
\end{split}
\end{eqnarray}
in which $b=1-2a\Omega_0$, $\Phi_0=-\frac{GM}{R}$ and $\lambda=-\frac{GM}{R^2}$. $a$ is the angular momentum per mass and $\Omega_0$ the local angular velocity of the Kerr spacetime. The authors set $\Phi_0=0$ for the first order of approximations. This does not works because $\lambda$ is related to $\Phi_0$ through $\gamma=-R\Phi_0$. Putting one of them equal to zero forces the other one to vanish also. Evidently, in the first order of approximations we must keep both $\Phi_0$ and $\gamma$ and the metric must be written as follows instead of (\ref{eq60}):
\begin{eqnarray}\label{eq61}
\begin{split}
ds^2 \approx & dt^2 - [dx^2+dy^2+dz^2]\;\;\; 0-th\; order\; of\; approximation ,\\
ds^2 \approx & [1 +2b(\Phi_0+\gamma z)]dt^2 - [1 -2(\Phi_0+\gamma z)][dx^2+dy^2+dz^2]\;\;\; first\; order\; of\; approximation.
\end{split}
\end{eqnarray}
Taking the above comment and the metric (\ref{eq61}) into account, we find $\gamma_0=b\Phi_0,\; \lambda_0=b\gamma=-bR\Phi_0,\; \gamma_1=\Phi_0,\; \lambda_1=\gamma=-bR\Phi_0$ and arrive at the following relations for the e\emph{lectromagnetic Casimir energy density} and the mode frequencies:
\begin{eqnarray}\label{eq62}
\begin{split}
\omega &=\omega_0[1+b\Phi_0+b\frac{\gamma l_P}{2}] ,\\
<0|T_{00}^{Maxwell}|0> &=-\frac{\pi^2}{720 l^4}[1+b\Phi_0+b\frac{\gamma l_P}{2}].
\end{split}
\end{eqnarray}
The scalar field Casimir energy density inside the apparatus which has been sketched in eq.(32) in \cite{Bezerra} must also be corrected as follows:
\begin{eqnarray}\label{eq63}
\begin{split}
\omega &=\omega_0[1+b\Phi_0+b\frac{\gamma l_P}{2}] , \\
<0|T_{00}^{scalar}|0> &=-\frac{\pi^2}{1440 l^4}[1+b\Phi_0+b\frac{\gamma l_P}{2}].
\end{split}
\end{eqnarray}
\emph{Far field limit of the Schwarzschild spacetime} is also covered by setting $b=1$

\subsection{The Fermi spacetime}
The Fermi spacetime is described by the following metric
\begin{eqnarray}\label{eq64}
ds^2=(1+2a z)dt^2-dx^2-dy^2-dz^2
\end{eqnarray}
The importance of this metric is that it is traditionally recognized as the spacetime of a static accelerating observer near the surface of the source of a constant gravitational field \cite{Misner}. Comparison of this metric to the general metric in (\ref{eq07}) gives $\gamma_0=0, \; \lambda_0=a, \gamma_1=\lambda_1=0$:
\begin{eqnarray}\label{eq65}
\begin{split}
\omega &=\omega_0[1+\frac{a l_P}{2}] ,\\
<0|T_{00}^{Maxwell}|0> &=-\frac{\pi^2}{720 l^4}[1+\frac{a l_P}{2}], \\
<0|T_{00}^{scalar}|0> &=-\frac{\pi^2}{1440 l^4}[1+\frac{a l_P}{2}].
\end{split}
\end{eqnarray}
The above energy densities are exactly the results (5.2) in \cite{Bimonte}, (3.4) in \cite{Napolitano} and (5.4) in \cite{Esposito}.
\subsection{The Ho\v{r}ava-Lifshitz gravity with a cosmological constant}
The Ho\v{r}ava-Lifshitz (HL) gravity is a renormalizable theory of gravity that is invariant under the Lifshitz scaling transformation $\textbf{x}\rightarrow b\textbf{x}, t\rightarrow b^z t$. This transformations manifestly break the space and time covariance. The anisotropy between space and time, in turn, may affects the Casimir effect as well. It is interesting to investigate the vacuum characteristics of the theory. Recently, the effect of the HL theory on the Casimir energy of the apparatus has been studied in \cite{Ferrari}. The authors recommended to set a constraint on spacetime anisotropies in such a way that the Casimir energy modifications remain within the experimental bounds. Recently in \cite{Muniz1} , \cite{Muniz2} the same problem considered in curved spacetime in the context of a spherical symmetric solution of the HL theory. Finite temperature Casimir energy in spacetime (\ref{eq07}) has been analysed by the author in \cite{BorzooNew} and the following weak field limit for the HL theory has been calculated:
\begin{eqnarray}\label{eq66}
\begin{split}
ds^2=\{1 &+[2\frac{\widehat{M}}{R}+\frac{3\widehat{M}^2}{2R^2}-\frac{\widehat{M}^2}{2\widehat{\omega} R^4}]
+[-2\frac{\widehat{M}}{R^2}-\frac{3\widehat{M}^2}{2R^3}+\frac{\widehat{M}^2}{2\widehat{\omega} R^5}]z\} dt^2 \\
&-\{1+[-\frac{\widehat{M}}{R}+\frac{\widehat{M}^2}{4R^2}+\frac{\widehat{M}^2}{4\widehat{\omega}R^4}]
+[\frac{\widehat{M}}{R^2}-\frac{\widehat{M}^2}{4R^3}-\frac{\widehat{M}^2}{4\widehat{\omega}R^5}]z \} (dx^2+dy^2+dz^2)
\end{split}
\end{eqnarray}
where $\widehat{M}=M(1+\frac{\Lambda}{\omega})$. This spacetime is the weak field limit of the Park's spherical symmetric solution to the IR limit of the HL theory \emph{in the presence of a cosmological constant\cite{Park}} \cite{Muniz2}. $\Lambda$ may have the same role as the cosmological constant but not necessarily being a small parameter and $\omega$ is a constant frequently used to regulate the UV limit of the HL theory. Comparison between (\ref{eq66}) and the general metric (\ref{eq07}) shows:
\begin{eqnarray}\label{eq67}
\begin{split}
\gamma_0\approx \frac{M(1+\frac{\Lambda}{\omega})}{R}, \; \lambda_0\approx -\frac{M(1+\frac{\Lambda}{\omega})}{R^2}, \gamma_1\approx\frac{M(1+\frac{\Lambda}{\omega})}{2R},\lambda_1\approx -\frac{M(1+\frac{\Lambda}{\omega})}{2R^2},
\end{split}
\end{eqnarray}
Another spacetime, which is a solution to the HL theory \emph{without cosmological constant} is the Kehagias-Sfetsos (KS) solution. This spacetime has been discussed in \cite{Muniz1} to obtain the energy density of the apparatus and is as follows:
\begin{eqnarray}\label{eq68}
ds^2=f_{KS}dt^2-f^{-1}_{KS}d\rho^2-\rho^2d\Omega^2
\end{eqnarray}
where $f_{KS}=1+\omega\rho^2(1-\sqrt(1+\frac{4M}{\omega\rho^3}))$. $\rho$ is a radial coordinate and $\omega$ is the free parameter of the HL theory. Putting $\Lambda=0$ in the Park's solution recover the KS solution and so the coefficients in equation (\ref{eq67}) are also valid for the KS solution.

\section{Conclusion}
We analysed the energy density of a Casimir apparatus consisting of two nearby conducting parallel plates in a general weak gravitational field. The metric in the equation (\ref{eq07}) denotes the deviation of the weak gravitational field from flat spacetime inside the apparatus. We transformed the metric (\ref{eq07}) through a gauge transformation into the Fermi metric and then  solved the Klein-Gordon equation exactly. The mode frequencies were found for the scalar field inside the apparatus for both Neumann and Dirichlet boundary conditions in terms of the weak gravitational field parameters $\gamma_0,\gamma_1,\lambda_0,\lambda_1$. This result was shown to be valid also for the electromagnetic field in section V. The energy density of the apparatus was found for both scalar and electromagnetic fields in terms of the weak field parameters. Some examples of weak gravitational fields were analysed in section VII. Specially the electromagnetic energy density and mode frequencies in the far field limit of the Kerr spacetime in its equatorial plane were obtained. The weak field limit of the Ho\v{r}ava-Lifshitz gravity with a cosmological constant was also investigated and the weak field parameters were sketched. Consistency of the results with the literature was checked by considering the Fermi metric.

\pagebreak
\appendix
\section{Asymptotic form of the wave function}
In this section we find an explicit and simple asymptotic form for the wave function. As is apparent we need to have an asymptotic expansion with both argument and first parameter being large. Whittaker functions have such an expansion in terms of Airy functions \cite{Olver}:
\begin{eqnarray}\label{A1}
W_{\kappa,\mu}(4\kappa x)=2^{4/3}\sqrt{\pi}\kappa^{\kappa+1/6}(\frac{x\zeta}{x-1})^{\frac{1}{4}}\{Ai[(4\kappa)^{\frac{2}{3}}\zeta]\sum_{n=0}^{\infty}\frac{A_n(\zeta)}{(4\kappa)^{2n}}+\frac{Ai^{'}[(4\kappa)^{\frac{2}{3}}\zeta]}{(4\kappa)^{\frac{4}{3}}}\sum_{n=0}^{\infty}\frac{B_n(\zeta)}{(4\kappa)^{2n}}\}
\end{eqnarray}
where $\zeta$ is defined as
\begin{eqnarray}\label{A3}
\frac{4}{3}(-\zeta)^{\frac{3}{2}}=cos^{-1}(\sqrt{x})-\sqrt{x-x^2}
\end{eqnarray}
and in our case $4\kappa x=\frac{k_\perp}{\lambda}g_{00}$ and so
\begin{eqnarray}\label{A2}
x=\frac{k_\perp^2}{\omega^2}g_{00}=\frac{k_\perp^2}{\omega^2}(1+2\lambda z) \;\;\ , \;\;\ x<1 \;\;\ or \;\;\ S\equiv\frac{\omega^2}{g_{00}}-k_\perp^2 >0
\end{eqnarray}
 If $\kappa \rightarrow \infty$ then both the argument and the first parameter go to infinity. As $\kappa \propto \lambda ^{-1}$ the second term in the bracket is of order $\lambda^{\frac{4}{3}}$ and must be ignored to stay within the first order of expansion in $\lambda$. Also the summation in the first term reduces to only $n=0$ term and $A_0(\zeta)=constant$.
\begin{eqnarray}\label{A4}
W_{\kappa,\mu}(4\kappa x)=2^{4/3}\sqrt{\pi}\kappa^{\kappa+1/6}(\frac{x\zeta}{x-1})^{\frac{1}{4}} Ai[(4\kappa)^{\frac{2}{3}}\zeta]
\end{eqnarray}
The argument of the above Airy function can be written as
\begin{eqnarray}\label{A5}
(4\kappa)^{\frac{2}{3}}\zeta=-\{3\kappa[cos^{-1}(\sqrt{x})-\sqrt{x-x^2}]\}^{\frac{2}{3}}\equiv-v(x) \;\;\ , \;\;\ v\rightarrow\infty
\end{eqnarray}
The Airy function with large argument is as follows (section 9.7 from \cite{NIST})
\begin{eqnarray}\label{A6}
Ai(-v)\sim v^{-\frac{1}{4}} \{cos(\frac{2}{3}v^{\frac{3}{2}}-\frac{\pi}{4})\sum_{n=0}^\infty(-1)^n\frac{a_{2n}}{(\frac{2}{3}v^{\frac{3}{2}})^{2n}}+sin(\frac{2}{3}v^{\frac{3}{2}}-\frac{\pi}{4})\sum_{n=0}^\infty(-1)^n\frac{a_{2n+1}}{(\frac{2}{3}v^{\frac{3}{2}})^{2n+1}}\}
\end{eqnarray}
As $v^{\frac{3}{2}}=O(\lambda^{-1})$, the $sin()$-part and $n>0$ in the $cos()$-part must be ignored also.
The observation that
\begin{eqnarray}\label{A7}
cos^{-1}(\sqrt{x})-\sqrt{x-x^2} =-\int \sqrt{\frac{1-x}{x}}dx=-\frac{1}{k_\perp}\int \sqrt{S}dx=-\frac{1}{2\kappa}\int \sqrt{S}dz,
\end{eqnarray}
and
\begin{eqnarray}\label{A8}
(\frac{x\zeta}{x-1})^{\frac{1}{4}} \propto v^{\frac{1}{4}}S^{-\frac{1}{4}}\;\;\;\;\;\;\;\;\;\;\;\;\;\;\;\;\;\;\;\;\;\;\;\;\;\;\;\;\;\;\;\;\;\;\;\,
\end{eqnarray}
moves the situation forward significantly:
\begin{eqnarray}\label{A9}
W_{\kappa,\mu} \propto S^{-\frac{1}{4}} cos(\int \sqrt{S}dz+ \phi_0) \;\;\ , \;\;\ \phi_0 =\phi_0(\kappa, \omega)
\end{eqnarray}
A same process can be applied to the other Whittaker's function $M_{\kappa,\mu}$ except that according to \cite{Erdélyi} instead of the a argument $(4\kappa)^{\frac{2}{3}}\zeta$ in (\ref{A5}) we have $(4\kappa)^{\frac{2}{3}}exp(\pm \frac{2\pi}{3}) \zeta $. This difference changes the above process as follows:
\begin{eqnarray}\label{A10}
\zeta\rightarrow e^{\pm \frac{2\pi}{3}} \zeta  \;\;\;\ , \;\;\;\ v\rightarrow e^{\pm \frac{2\pi}{3}} v  \;\;\;\ ,\;\;\;\ \frac{2}{3}v^{\frac{3}{2}} \rightarrow \frac{2}{3}v^{\frac{3}{2}} e^{\pm \pi i}=- \frac{2}{3}v^{\frac{3}{2}}
\end{eqnarray}
\begin{eqnarray}\label{A11}
M_{\kappa,\mu} \propto S^{-\frac{1}{4}} cos(\int \sqrt{S}dz+ \phi_1) \;\;\ , \;\;\ \phi_1 =\phi_1(\kappa, \omega)
\end{eqnarray}
As a result the total wave function in (\ref{eq22}) when $\lambda \rightarrow 0$ reads
\begin{eqnarray}\label{A12}
\phi_{\kappa}(u)=C_0(\omega,k_\perp) (g_{00}S(z))^{-\frac{1}{4}} sin(\int \sqrt{S}dz+\phi_2)  e^{-i\omega t-ik_xx -ik_yy}\;\;\;\;\;\;\;\;\
\end{eqnarray}
where $\phi_2=\phi_2(\omega,k_\perp, A,B,\phi_0,\phi_1)$ and $A,B$ came from (\ref{eq22}) and the following relation has been used:
\begin{eqnarray}\label{A13}
A cos(\int \sqrt{S}dz+\phi_0)+B cos(\int \sqrt{S}dz+\phi_1)=C sin(\int \sqrt{S}dz+\phi_2)
\end{eqnarray}
\section{Computation of the components of the energy-momentum tensor}
In this section we find $T^{ij }$. Other components will be find by similar techniques. We do the computations in the Fermi spacetime i.e. $g_{00}=1+2\lambda z, g_{0i}=0, g_{ij}=-\delta_{ij}$. Greece indices run from 0 to 3 and Latins from 1 to 3. From (\ref{eq36}) we have
\begin{eqnarray}\label{B1}
T^{ij }_{Maxwell} =\frac{1}{4}g^{ij}F^{\lambda\theta}F_{\lambda\theta}-F^{i\theta}F_{\; \theta}^{j}.
\end{eqnarray}
The first term has the following form:
\begin{eqnarray}\label{B2}
\begin{split}
F^{\lambda\theta}F_{\lambda\theta} &=2F^{0i}F_{oi}+  F^{ij}F_{ij}, \\
F^{0i}F_{oi} &=g^{00}E^i E_i, \\
F^{ij}F_{ij} &=g^{\mu i}g^{\nu j} F_{\mu\nu}\varepsilon_{ijk} B^k=...=-2B^2.
\end{split}
\end{eqnarray}
where we have used $\varepsilon_{ijk}\varepsilon_{ijl}=2\delta_{kl}$. The second term in (\ref{B1}) simplifies as follows:
\begin{eqnarray}\label{B3}
\begin{split}
F^{i\theta}F_{\; \theta}^{j} &=F^{i0}F_{\;0}^{j}+  F^{im}F^{j}_{\;m}, \\
&F^{i0}=g_{00}E^i, \; F_{\;0}^{j}=-E^j, \\
&F^{im}=g^{il}g^{mk}F_{lk}=g^{il}g^{mk}\varepsilon{lkn}B^n, \; F^j_{\; m}=g^{aj}F_{am}=g^{aj}\varepsilon{amb}B^b, \\
&F^{im}F^j_{\;m}=-g^{il}g^{aj}\varepsilon_{mab}\varepsilon_{mln}B^bB^n \\
&\;\;\;\;\;\;\;\;\;\;\;=-g^{il}g^{aj}[\delta_{al}\delta_{bn}-\delta_{an}\delta_{bl}]B^bB^n \\
&\;\;\;\;\;\;\;\;\;\;\;=...=-g^{ij}B^2+B^iB^j.
\end{split}
\end{eqnarray}
Based on (\ref{B1})-(\ref{B3}) we find
\begin{eqnarray}\label{B4}
T^{ij }_{Maxwell} =\frac{1}{2}g^{ij}(g^{00}E^2+B^2)-g^{00}E^iE^j-B^iB^j.
\end{eqnarray}
After lowering the indices we have finally:
\begin{eqnarray}\label{B5}
T_{ij }^{Maxwell} =\frac{1}{2}g_{ij}(g^{00}E^2+B^2)-g^{00}E_iE_j-B_iB_j.
\end{eqnarray}

Now we find $T_{0i}$. From (\ref{eq36}) we have
\begin{eqnarray}\label{B6}
T^{00}_{Maxwell} =\frac{1}{4}g^{00}F^{\lambda\theta}F_{\lambda\theta}-F^{0\theta}F_{\; \theta}^{0},
\end{eqnarray}
in which the first term already has been obtained in (\ref{B2}). The second term has been also obtained in (\ref{B3}) and after lowering the indices again we have
\begin{eqnarray}\label{B7}
T_{00}^{Maxwell} =-\frac{1}{2}(E^2+g_{00}B^2).
\end{eqnarray}

Now we find $T_{0i}$. From (\ref{eq36}) we have
\begin{eqnarray}\label{B8}
\begin{split}
T^{0i}_{Maxwell} &=\frac{1}{4}g^{0i}F^{\lambda\theta}F_{\lambda\theta}-F^{0\theta}F_{\; \theta}^{i} \\
&=-F^{0j}F_{\; j}^{i}=-g_{00}E^jg^{mi}F_{mj}=-g_{00}g^{mi}\varepsilon_{mjk}E^jB^k=-g_{00}g^{mi}(\textbf{E}\times \textbf{B})_m,
\end{split}
\end{eqnarray}
from which, after lowering the indices again, we find
\begin{eqnarray}\label{B9}
\begin{split}
T_{0i}^{Maxwell} =-(\textbf{E}\times \textbf{B})_i.
\end{split}
\end{eqnarray}
\section *{Acknowledgments}
The author would like to thank University of Tehran for supporting this project under the grants provided by the research council.
\pagebreak

\end{document}